\documentclass[showpacs,superscriptaddress,prl,twocolumn]{revtex4}

\usepackage{amsmath}
\usepackage{graphicx}
\usepackage{bm}
\usepackage{amsfonts}
\usepackage{amssymb}

\begin{document}

\title{Coherent backscattering of Bose-Einstein condensates
  in two-dimensional disorder potentials}

\author{Michael Hartung}
\affiliation{Institut f\"ur Theoretische Physik,
Universit\"at Regensburg, 93040 Regensburg, Germany}
\author{Thomas Wellens}
\affiliation{Physikalisches Institut, Albert-Ludwigs Universit\"at
  Freiburg, 79104 Freiburg, Germany} 
\author{Cord A.~M\"uller}
\affiliation{Physikalisches Institut, Universit\"at Bayreuth, 95440 Bayreuth,
  Germany} 
\author{Klaus Richter}
\author{Peter Schlagheck}
\affiliation{Institut f\"ur Theoretische Physik,
Universit\"at Regensburg, 93040 Regensburg, Germany}

\pacs{05.60.Gg; 03.75.Kk; 72.15.Rn}

\begin{abstract}

We study quantum transport of an interacting Bose-Einstein condensate
in a two-dimensional disorder potential. In the limit of vanishing atom-atom
interaction, a sharp cone in the angle-resolved density of the scattered 
matter wave is observed, arising from constructive interference between
amplitudes propagating along reversed scattering paths.
Weak interaction transforms this coherent backscattering peak into a
pronounced dip, indicating destructive instead of constructive interference.
We reproduce this result, obtained from the numerical integration of the 
Gross-\-Pi\-ta\-evskii equation, by a diagrammatic theory of weak localization
in presence of a nonlinearity.

\end{abstract}

\maketitle


The past years have witnessed an  increasing number of theoretical and
experimental research activities on the behaviour of ultracold atoms in
magnetic or optical disorder potentials
\cite{CleO05PRL,ForO05PRL,SchO05PRL,BilO08xxx,Roa08xxx,SanO07PRL,PauO05PRA,PauO07PRL,BilPav06EPJD,LugO07PRL,LyeO07PRA,CheO08PRA,DamO03PRL,FalO07PRL}.
A central aim in this context is the realization and unambiguous
identification of strong Anderson localization with Bose-Einstein
condensates, which was attempted by several experimental groups
\cite{CleO05PRL,ForO05PRL,SchO05PRL} with recent success
\cite{BilO08xxx,Roa08xxx}, and theoretically studied both
from the perspective of the expansion process of the condensate
\cite{SanO07PRL} as well as from the scattering perspective
\cite{PauO05PRA,PauO07PRL}.
Complementary studies were focused on localization properties of Bogoliubov
quasiparticles \cite{BilPav06EPJD,LugO07PRL}, on dipole oscillations in
presence of disorder \cite{LyeO07PRA,CheO08PRA}, as well as on the realization
of Bose glass phases \cite{DamO03PRL,FalO07PRL}.

The above-mentioned topics mainly refer to
processes that are essentially one-dimensional (1D) by nature.
Qualitatively new phenomena, however, do arise in two or three spatial
dimensions, due to the scenario of \emph{weak localization}.
The latter manifests in a slight reduction of the transmission probability of
an incident wave through a disordered region as compared to the classically
expected value, due to constructive interference between backscattered paths
and their time-reversed counterparts.
This interference phenomenon particularly leads to a cone-shaped enhancement
of the backscattering current in the direction reverse to the incident
beam, which was indeed observed \cite{VanLag85PRL} and theoretically analyzed
\cite{AkkWolMay86PRL} in light scattering processes from disordered media.
Related weak localization effects also arise in electronic mesoscopic physics,
leading to characteristic peaks in the magneto-resistance
\cite{AltO80PRB,Ber84PR}.

In this Letter, we investigate the phenomenon of coherent backscattering with
atomic Bose-Einstein condensates that propagate in presence of two-dimensional
(2D) disorder potentials.
An essential ingredient that comes into play here is the \emph{interaction}
between the atoms of the condensate. On the mean-field level, this is
accounted for by the nonlinear term in the Gross-Pitaevskii equation
describing the time evolution of the condensate wavefunction. 
Indeed, nonlinearities do also appear in scattering processes of light e.g.\ 
from a gas of cold atoms, due to the saturation of the intra-atomic transition
\cite{ChaO04PRE,ShaMulBuc05PRL,WelO06PRA}.
In this case, however, the saturation also leads to inelastic scattering
\cite{ShaMulBuc05PRL,WelO06PRA} and, in addition, the nonlinearity competes
with other dephasing mechanisms induced, e.g., by polarization phenomena
\cite{JonO00PRL} or thermal motion \cite{LabO06PRL}. 
The complementary process of atomic condensates scattering from optical random
potentials in the mean-field regime provides a cleaner situation where the
coherence of the atomic wavefunction remains well preserved in the presence of
the nonlinearity.
As we shall argue below, this leads to substantial modifications of the
coherent backscattering feature. In particular, the interaction turns
constructive into destructive interference, leading to a negative coherent
backscattering peak height. This is reminiscent of the weak antilocalization
effects due to spin-orbit interaction observed for mesoscopic magnetotransport
\cite{ZumO02PRL}.

The starting point of our investigation is the time-dependent 2D
Gross-Pitaevskii equation describing the mean-field dynamics of the condensate
in presence of the disorder potential $V(\vec{r})$ [$\vec{r} \equiv (x,y)$],
 \begin{eqnarray}
   i\hbar\frac{\partial}{\partial t} \psi(\vec{r},t) & = &
   \left(-\frac{\hbar^2}{2 m} \Delta + {V(\vec{r})} + \tilde{g}(x)
     |\psi(\vec{r},t)|^2 \right) \psi(\vec{r},t) \nonumber \\
      & & + \;{S(t)  \delta(x-x_0) \exp(-i \mu t / \hbar)} \, ,
      \label{Eq:GP}
\end{eqnarray}
where $S(t)$ denotes a source term simulating the coherent injection of matter
waves with chemical potential $\mu$ from an external reservoir onto the
scattering region \cite{PauO05PRA}.
In the numerical integration of Eq.~(\ref{Eq:GP}), $S(t)$ is adiabatically
increased from zero to a final value $S_0$ that corresponds to a fixed
incident current density $j_{\rm in}$.
Periodic boundary conditions are imposed on the transverse boundaries (in $y$
direction) of the numerical grid to ensure a homogeneous flow in absence of 
disorder, whereas absorbing boundary conditions applied at the edges of the
longitudinal ($x$) direction allow us to inhibit artificial backreflection of
outgoing waves with rather high accuracy \cite{Shi91PRB}.

\begin{figure}
  \includegraphics[width=\linewidth]{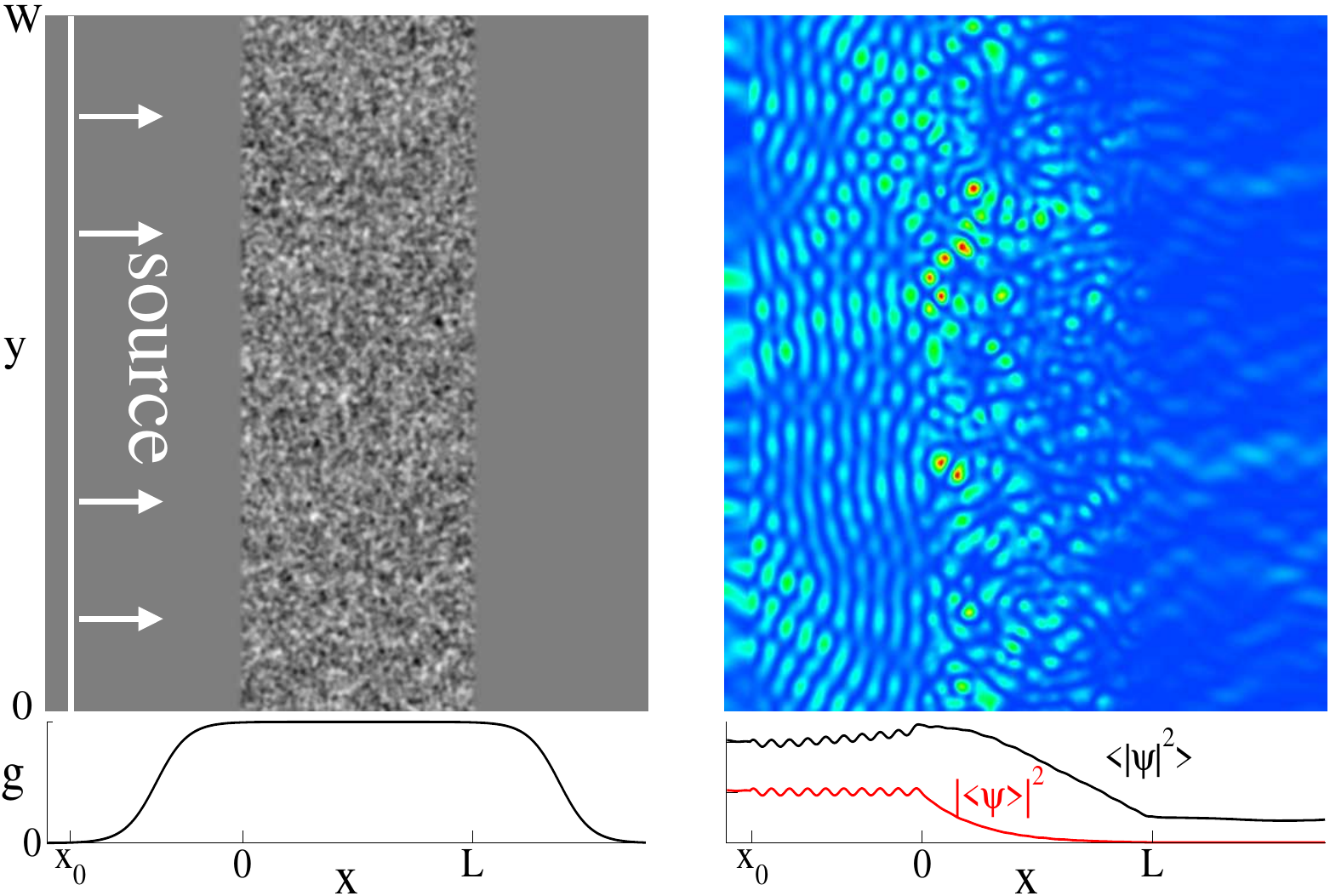}
  \caption{(Color online)
    Scattering geometry and stationary scattering state associated with a
    randomly generated disorder potential.
    The left-hand side displays $V(x,y)$ in a gray-scale plot and shows the
    spatial variation of the nonlinearity $g(x)$.
    The upper right panel shows the density of the corresponding scattering
    state that is populated through the numerical integration of the
    inhomogeneous Gross-Pitaevskii equation (\ref{Eq:GP}).
    The lower right panel shows the decay of the coherent mode
    $|\langle \psi \rangle|^2$ and the density 
    $\langle |\psi|^2 \rangle$ with $x$, averaged over $y$ for $\sim 10^3$ 
    randomly generated disorder configurations.
    Parameters: $kL = 40$, $kW = 120$, $k\sigma = 0.5$, $V_0 = 0.614 \mu$, 
    $g = 0.005$, $j_{\rm in} = \hbar k^3 / m$, with 
    $k \equiv \sqrt{2 m \mu} / \hbar$.
  }
  \label{Fig:Setup} 
\end{figure}

In Eq.~(\ref{Eq:GP}), the effective 2D interaction strength is written as 
$\tilde{g}(x) \equiv \hbar^2 g(x) /(2 m)$, with the dimensionless nonlinearity
parameter $g(x)$.
In presence of a harmonic confinement of the condensate in the third
spatial dimension with the oscillator length 
$a_\perp(x) \equiv \sqrt{\hbar/[m\omega_\perp(x)]}$, we have 
$g(x) = 4\sqrt{2\pi} a_s / a_\perp(x)$, where $a_s$ denotes the $s$-wave
scattering length of the atoms.
We assume that $g(x)$ is adiabatically ramped on and off in front of and behind
the disorder region, as shown in Fig.~\ref{Fig:Setup}.
Physically, this spatial variation of the nonlinearity, which is needed in
order to avoid nonlinear effects at the position of the source and the
absorbing boundaries, would correspond to a finite extent of the transverse
harmonic confinement into which the condensate is propagating. 
As for the disorder potential $V(\vec{r})$, we choose a Gaussian random process
characterized by a vanishing mean value $\langle V(\vec{r})\rangle=0$ and a
Gaussian correlation function
$\langle V(\vec{r})V(\vec{r}+\Delta\vec{r}) \rangle=
  V_0^2\textrm{e}^{-\Delta r^2/2\sigma^2}$ 
with correlation length $\sigma$.
We focus in the following on the parameters  $k \sigma = 0.5$, with 
$k \equiv \sqrt{2 m \mu} / \hbar$ the wavenumber of the incident beam, and
$V_0 / \mu = 0.614$.
The incident current density reads $j_{\rm in} = \hbar k |\psi_0|^2 / m$,
where we set $\psi_0 = k$ for the amplitude of the incident wave \cite{remark}.

\begin{figure}[t]
  \includegraphics[width=\linewidth]{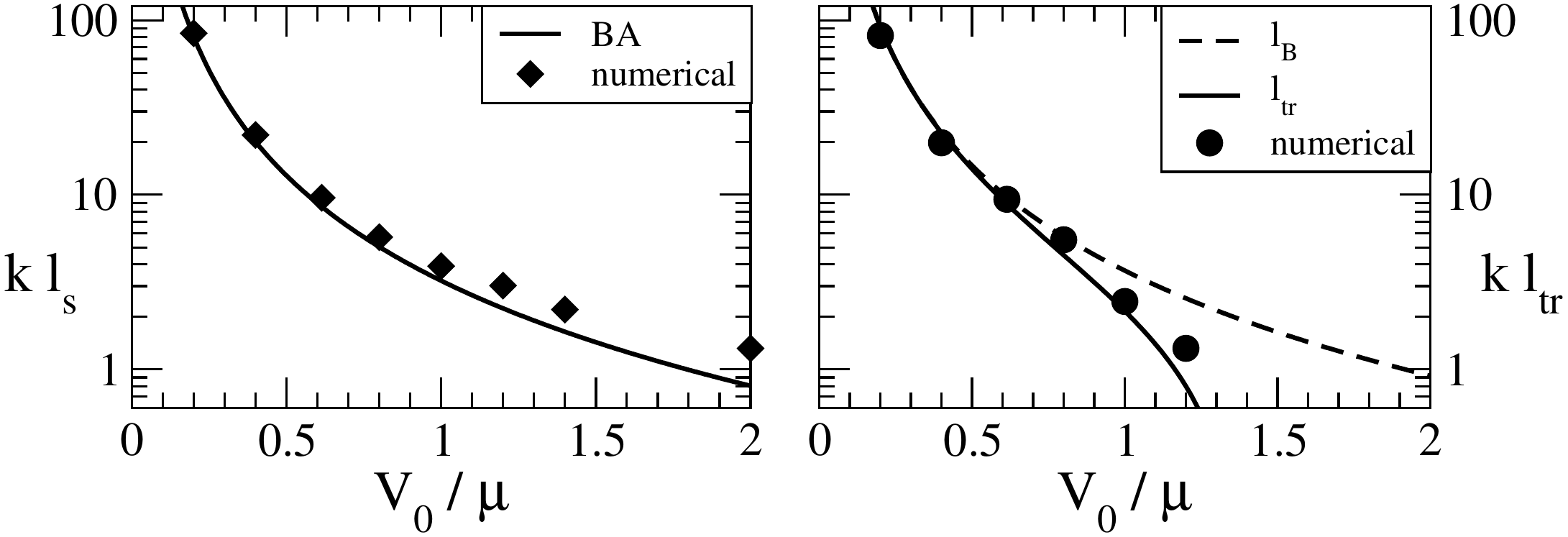}
 	\caption{Scattering mean free path $\ell_s$ (left panel) and transport mean
    free path $\ell_{tr}$ (right panel) in the disorder potential for $k
    \sigma = 0.5$ in absence of the nonlinearity. In the left panel,
    the numerically computed lengths are compared with the Born approximation
    (\ref{Eq:ls}) (solid line) and in the right panel with the Boltzmann mean
    free path (\ref{Eq:lB}) ($\ell_B$, dashed line) and the expression
    (\ref{Eq:lt}) that takes into account weak localization corrections
    ($\ell_{tr}$, solid line).
    We find $\ell_{tr} \simeq \ell_s$ for $V_0 < \mu$, which characterizes
    isotropic scattering. 
  }
	\label{Fig:meanfreepath}
\end{figure}

At the above values for $\sigma$ and $V_0$, scattering in the disorder region
is approximately isotropic.
This is quantitatively expressed by the equivalence of the two relevant length
scales that the disorder averages introduces for the transport process of the
condensate:
the scattering mean free path $\ell_s$, which describes the average decay of
the incident coherent mode inside the disorder region according to
$|\langle \psi(\vec{r}) \rangle|^2 \propto \exp(-x/\ell_s)$, 
and the transport mean free path $\ell_{\rm tr}$, which characterizes the
decay of the average density $\langle |\psi(\vec{r})|^2 \rangle$ 
(see Fig.~\ref{Fig:Setup}).
In absence of the nonlinearity, the scattering mean free path is in leading
order in $V_0$ given by the Born approximation
\begin{equation}
  (k\ell_s)^{-1} \simeq (\pi/2) (V_0/\mu)^2 (k \sigma)^2
  I_0(k^2 \sigma^2) \exp(-k^2 \sigma^2)
  \label{Eq:ls}
\end{equation}
where $I_j(\xi)$ is the modified Bessel function of order $j$.

The transport mean free path can be extracted from the linear decrease 
of $\langle |\psi(\vec{r})|^2 \rangle$ with $x$ according to
 $\langle |\psi(\vec{r})|^2 \rangle \propto (L + z_0 \ell_{tr} - x)$,
with $z_0=0.82$ in two spatial dimensions, and $L$ the
longitudinal extent of the disorder region.
In lowest order in $V_0$, $\ell_{\rm tr}$ is given by the Boltzmann transport
mean free path $\ell_B$ defined through
\begin{equation}
  \ell_s/\ell_B=1-I_1(k^2\sigma^2)/I_0(k^2\sigma^2) \, . \label{Eq:lB}
\end{equation}
Weak localization effects lead to logarithmic corrections that yield
for $k\ell_B \gg 1$ \cite{GorLarKhm79JETPL,KuhO05PRL}
\begin{equation}
  \ell_{tr} \simeq \ell_B [ 1- 2(k\ell_B)^{-1} \log(L/\ell_B) ] .
  \label{Eq:lt}
\end{equation}
As shown in Fig.~\ref{Fig:meanfreepath}, the expressions (\ref{Eq:ls}) and 
(\ref{Eq:lt}) are in good agreement with the numerically computed values of
$\ell_s$ and $\ell_{tr}$ for $V_0 < \mu$.
Specifically at $k\sigma=0.5$ and $V_0 / \mu = 0.614$, we find $k\ell_{s}
\simeq 9.61$ and $k\ell_{tr} \simeq 9.75$.

\begin{figure}[t]
   \includegraphics[width=\linewidth]{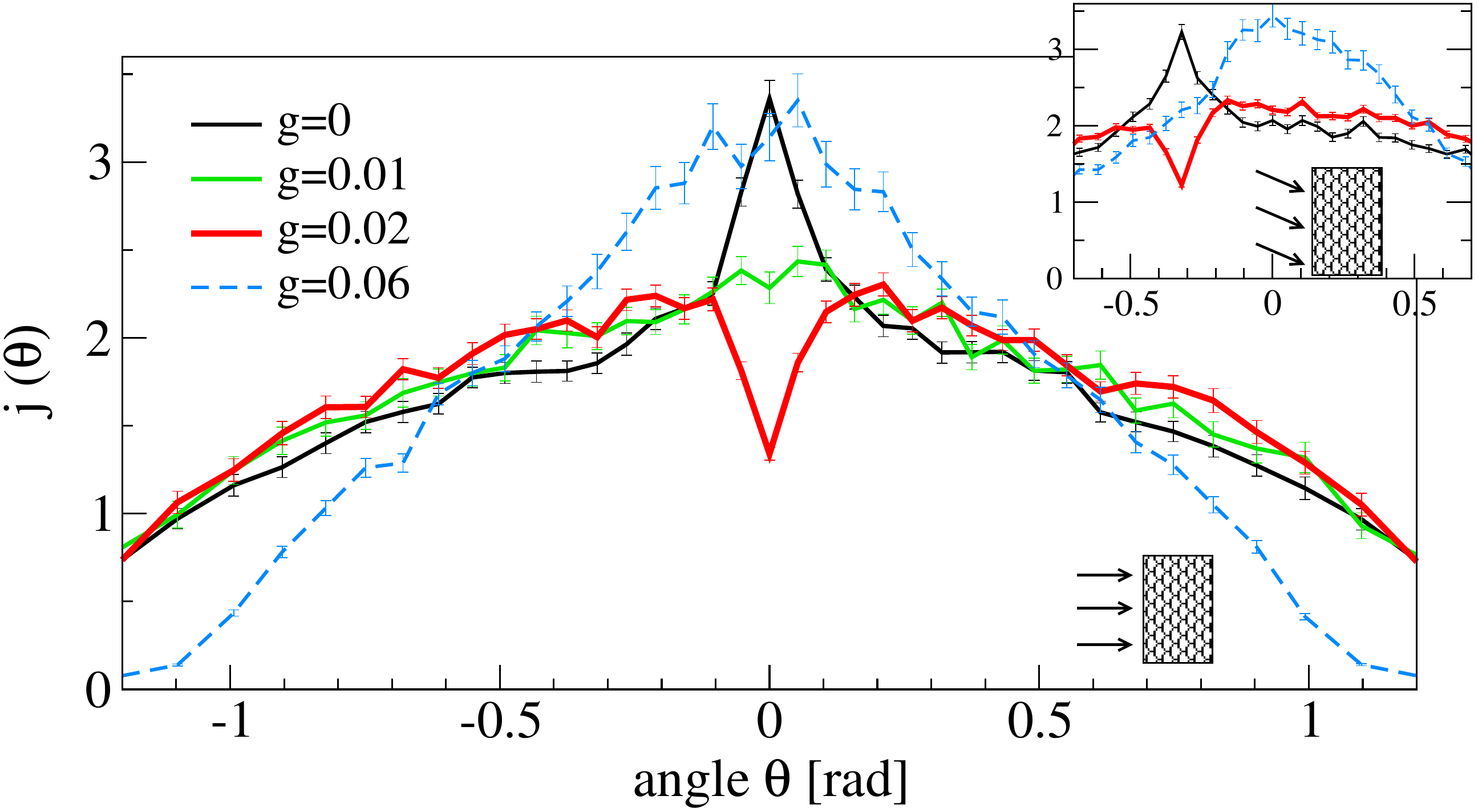}
   \caption{(Color online)
     Angle-resolved current density of backscattered atoms in absence
     and in presence of the nonlinearity, obtained from the average over 
     $\sim 10^3$ disorder configurations
     (parameters as in Fig.~\ref{Fig:Setup}; the error bars denote the
     statistical standard deviation).
     The coherent backscattering cone for $g=0$ (black line) is transformed
     into a pronounced dip for intermediate nonlinearities 
     ($g=0.02$, bold red line), and turns into a smooth peak structure 
     at larger values of $g$ ($g=0.06$, dashed blue line).
     The inset shows the angle-resolved current for the case of a
     \emph{tilted} incident beam where the source term in Eq.~(\ref{Eq:GP})
     populates the transverse eigenmode defined by the angle 
     $\theta_6 \simeq 0.32$. 
     In contrast to the smooth peak, the cone and dip structures are indeed
     found at the angle that corresponds to retro-reflection of the incident
     beam, which confirms that they both arise due to interference between
     reflected paths.
   }
   \label{Fig:backscattering} 
\end{figure}

The angle-resolved current in backward direction is numerically computed from
the decomposition of the reflected wave
$\psi_{\rm ref}(x,y) \equiv \psi(x,y) - \psi_0\exp(ikx)$ at fixed position $x$
close to $x_0$ [where $g(x)$ is negligibly small] into the transverse
eigenmodes $\chi_n(y) \sim \exp(i n \pi y / W)$, which support outgoing waves
into the directions with the angles
$\theta_n \equiv \textrm{arcsin}[2\pi n / (kW)]$.
Figure~\ref{Fig:backscattering} shows the average angular density $j(\theta)$
of the backscattered current, which is normalized such that $\int_0^{2\pi}
j(\theta) d\theta = 2\pi$.
In the linear case ($g=0$), we encounter the well-known cone structure
at $\theta=0$, which is a characteristic signature of weak localization
\cite{VanLag85PRL,AkkWolMay86PRL}.
Rather small values of $g\sim 0.02$ corresponding to
$\tilde{g} |\psi(\vec{r})|^2 \sim 10^{-2} \mu$,
are sufficient to substantially modify this cone-shaped peak.
Most interestingly, it is not washed out by the nonlinearity, but transformed
into a \emph{dip} that roughly has the same shape as the peak at $g=0$.
This indicates that the underlying interference phenomenon between reflected
scattering paths is still effective at finite $g$, but has turned from
constructive to destructive.

The occurence of a dip in the backscattered current is confirmed
by calculations based on the diagrammatic approach for weak localization in
presence of a nonlinearity \cite{WelO06PRA,WelGre08PRL}. 
Assuming the realization of a stationary scattering state, the average density
$\left<|\psi(\vec{r})|^2\right>$ is expressed in terms of ladder diagrams,
which amounts to neglecting interference, and thus describing wave transport
as a classical random walk.
This assumption is valid approximately  for a dilute medium, i.e.\ for 
$k\ell\gg 1$ with $\ell \equiv \ell_B\simeq \ell_{tr}\simeq \ell_s$ (for
isotropic scattering).
Furthermore, we assume the condition $g^2|\psi_0/ k|^4 k\ell \ll 1$ under
which scattering from the fluctuations $\tilde{g}|\psi(\vec{r})|^2$ of the
nonlinear refractive index is negligible compared to scattering from the
disorder potential $V(\vec{r})$ \cite{SpiZyu00PRL}.
Therefore, the average density $\left<|\psi(\vec{r})|^2\right>$ 
remains approximately  unaffected by the nonlinearity,
and thus is well described by linear transport theory.
From the average density, the flux backscattered in
direction $\theta=0$ results as 
$j_{\rm L}(0)=\int_0^L dx \exp(-x/\ell) \left<|\psi(x)|^2\right>/(\ell|\psi_0|^2)$.

In a second step, the coherent backscattering peak is calculated by means of
crossed (Cooperon) diagrams, describing interference between reversed
scattering paths.
Following the diagrammatic approach presented in Ref.~\cite{WelGre08PRL}, we
obtain the height of the coherent backscattering peak from the transport 
equations
\begin{eqnarray}
  C_c(x) & = & |\psi_0|^2e^{-\hat{x}/\ell}\left(1+\frac{i}{k}\int_{x_0}^x dx'
    g(x')C_1(x')\right), \label{cc}\\
  C_1(x) & = & \int_0^L
  \frac{dx'}{\pi\ell}\left[K_0\left(\left|\frac{\hat{x}-x'}{\ell} \right|\right)
    \Bigl(C_1(x')+C_c(x')\Bigr)+\right.\nonumber\\
  & & +\frac{i}{k}K_1\left(\left|\frac{\hat{x}-x'}{\ell}\right|\right)
  \langle|\psi(x')|^2\rangle\times\\
  & & \times\left.\int_{\min(x,x')}^{\max(x,x')}dx''g(x'')
    \Bigl(C_1(x'')+C_c(x'')\Bigr)\right] \nonumber \label{c1}
\end{eqnarray}
for the ``Cooperon intensity'' $C_1(x)$ and the ``coherent Cooperon
intensity'' $C_c(x)$, with $\hat{x}\equiv\max(x,0)$ and $K_{0,1}$ the modified
Bessel functions of the second kind.
The contribution to the flux scattered in backward direction then results as
\begin{eqnarray}
  j_{\rm C}(0) & = & {\rm Re} \int_{0}^L
  \frac{dx}{\ell|\psi_0|^2}e^{-x/\ell}\biggl(C_1(x)+\biggr.\nonumber\\
  & & \left.+ \frac{i}{k}\langle|\psi(x)|^2\rangle\int_{x_0}^x dx'
    g(x')C_1(x')\right). \label{jc}
\end{eqnarray}
Note that nonlinear processes also occur for $x_0<x<0$
where $V({\bf r})=0$ but $g(x)>0$ (see Fig.~\ref{Fig:Setup}). 
Hence, the cone height $j_{\rm C}(0)$ --- in contrast to the background
intensity $j_{\rm L}(0)$ --- explicitly depends on the spatial extent of the
nonlinearity region in front of the disorder potential, and can therefore be
tuned through the ramp-up of $g(x)$.

\begin{figure}[t]
  \includegraphics[width=0.7\linewidth]{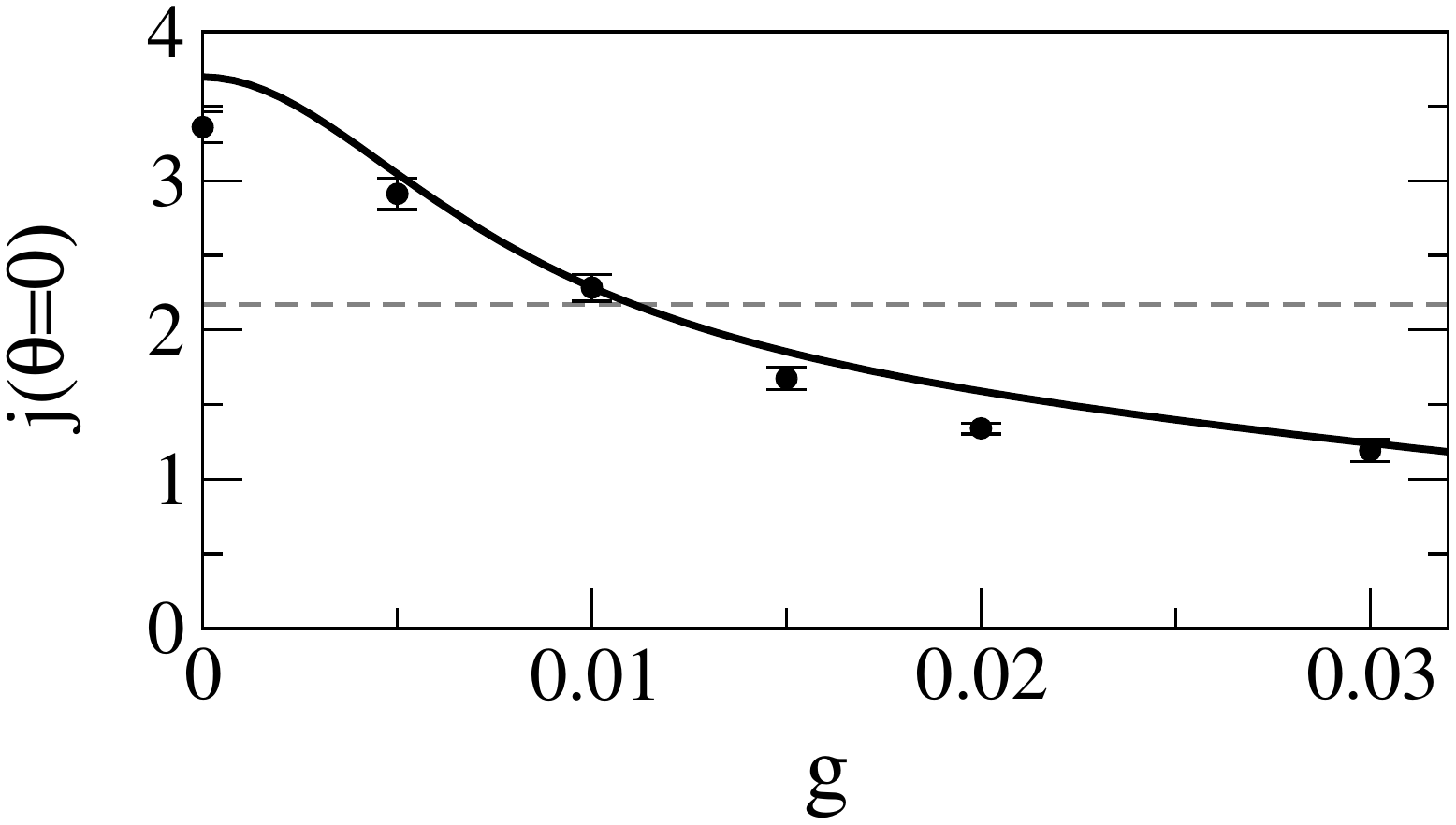}
  \caption{ 
    Backscattered current at $\theta=0$ as a function of the
    nonlinearity $g$ (parameters as in Fig.~\ref{Fig:Setup}), obtained from
    the numerical simulation (symbols) and from the diagrammatic theory,
    Eqs.~(\ref{cc}-\ref{jc}) (solid line). 
    The horizontal dashed line indicates the diffuse background intensity
    $j_{\rm L}(0)$.
    Negative cone heights $j_{\rm C}(0)<0$ leading to a dip in the 
    angle-resolved current density $j(\theta)$ appear for $g > 0.01$.
  }
  \label{Fig:comparison} 
\end{figure}

In the absence of nonlinearity, the above equations reduce to linear transport
theory in a two-dimensional slab, yielding
$\left.C_1(x)\right|_{g=0}=\left<|\psi(x)|^2\right>-|\psi_0|^2\exp(-x/\ell)$.
 We then obtain $\left.j_{\rm C}(0)\right|_{g=0}=j_{\rm L}(0)-1/2$, which expresses
reciprocity symmetry, i.e.\ the equality of reversed path amplitudes
(the term $1/2$ describes single scattering).
For $g\neq 0$, however, the nonlinearity turns $C_1(x)$ into a complex
quantity, as evident from the terms proportional to $ig$ in
Eqs.~(\ref{cc}-\ref{jc}).
This indicates an effective \emph{phase difference} between the reversed
scattering paths.
Consequently, the backscattered current $j_{\rm C}(0)$ is expected to decrease
with increasing nonlinearity, and may even become negative if this phase
difference is sufficiently large.

This latter situation is indeed encountered if the set of equations
(\ref{cc}-\ref{jc}) is numerically solved for the system parameters under
consideration.
As shown in Fig.~\ref{Fig:comparison}, the total flux $j_{\rm L}(0)+j_{\rm C}(0)$
resulting from Eqs.~(\ref{cc}-\ref{jc}) (solid line) agrees rather well with
the average value for $j(0)$ obtained from the numerical simulation (symbols).
Discrepancies are attributed to weak localization corrections in the
\emph{background} intensity, which would specifically lead to a reduction of 
the backscattered flux at $g=0$, and to an additional contribution to the
Cooperon intensity, termed $C_2(x)$ in Ref.~\cite{WelGre08PRL}, which was
neglected in the derivation of the above transport equations.
Details on these additional ingredients will be presented elsewhere.

At larger nonlinearities, $g \gtrsim 0.03$, the numerical propagation of the
inhomogeneous Gross-Pitaevskii equation (\ref{Eq:GP}) does not converge to a
stationary scattering state, but leads to a permanently time-dependent
behaviour of $\psi(\vec{r},t)$, as predicted in 
Refs.~\cite{SpiZyu00PRL} and encountered also in the
transport of condensates through 1D disorder potentials \cite{PauO05PRA}.
In this regime, the average backscattered current again displays a peak around
$\theta=0$;
this peak is, however, comparatively broad and does not arise from a
coherent backscattering phenomenon.
This becomes obvious if we inject the incident wave with a \emph{finite angle}
$\theta_6\simeq 0.32$ (corresponding to the transverse eigenmode $\chi_6$)
onto the disorder region.
While the cone and dip structures at $g=0$ and $0.02$ appear, as shown in the
inset of Fig.~\ref{Fig:backscattering}, at the expected angle of coherent
backscattering, corresponding to retro-reflection of the incident beam, the
broad peak at $g=0.06$ is not affected in this way.

In conclusion, the presence of a small nonlinearity reverts the scenario of
weak localization and gives rise to a cone-shaped dip, instead of a peak, in
the angle-resolved backscattered current density.
This phenomenon appears to be rather robust; it is numerically encountered
also for disorder potentials with longer correlation lengths $\sigma$
giving rise to anisotropic scattering, and we expect its manifestation also in
three spatial dimensions (as predicted by the diagrammatic theory) as well as
for speckle disorder where diagrammatic approaches would have to be based on
the treatment of Ref.~\cite{KuhO05PRL}.
We therefore believe that the effect would be measurable, for a reasonably
large range of parameters, in state-of-the-art transport experiments with
coherent Bose-Einstein condensates in well-controlled disorder potentials.

We thank A.~Buchleitner, D.~Delande, B. Gr\'emaud, R.~Kuhn and T.~Paul for
inspiring discussions. Funding through DFG (Forschergruppe 760) and
Bayerisches Elitef\"orderungsgesetz is gratefully acknowledged.


\begin{thebibliography}{10}
\bibitem{remark}
A different choice of $\psi_0$ would amount to rescaling $g$.
\end{thebibliography}
\end{document}